\documentclass[conference]{IEEEtran}

\IEEEoverridecommandlockouts

\usepackage{alltt}

\usepackage{tikz}
\usepackage{upquote}
\usetikzlibrary{decorations.pathreplacing}

\usepackage{graphicx}
\usepackage{comment}
\usepackage{subfig}
\usepackage{multirow}
\usepackage[normalem]{ulem} 
\usepackage{amstext} 
\usepackage{pifont}
\usepackage{MathDefs}
\usepackage{docmute}
\usepackage{circuitikz}
\usepackage{amsmath,amsfonts,amssymb,amsthm}
\usepackage{hyperref}
\hypersetup{
  colorlinks   = true, 
  urlcolor     = blue, 
  linkcolor    = blue, 
  citecolor   = red 
}
\usepackage{nomencl}
\makenomenclature
\usepackage{color, soul} 

\usepackage{tcolorbox} 
\newtcolorbox{Educ}[1]{
 title=#1,
  beamer, 
  colback=xlightblue,
  colframe=blue!30,
  fonttitle=\bfseries,
  left=1mm,
  right=1mm,
  top=1mm,
  bottom=1mm,
  middle=1mm,
  breakable,
}

\usepackage{url}
\usepackage{chapterbib}

\newtheorem{remark}{Remark}

\usepackage{algorithm}
\usepackage{algpseudocode} 

\usepackage{tabularx}
\newcolumntype{L}[1]{>{\raggedright\let\newline\\\arraybackslash\hspace{0pt}}m{#1}}
\newcolumntype{C}[1]{>{\centering\let\newline\\\arraybackslash\hspace{0pt}}m{#1}}
\newcolumntype{R}[1]{>{\raggedleft\let\newline\\\arraybackslash\hspace{0pt}}m{#1}}

\def\BibTeX{{\rm B\kern-.05em{\sc i\kern-.025em b}\kern-.08em
    T\kern-.1667em\lower.7ex\hbox{E}\kern-.125emX}}
\begin{document}

\title{Performance Analysis of Empirical Open-Circuit Voltage Modeling in Lithium Ion Batteries, \\Part-3: Experimental Results}

\author{\IEEEauthorblockN{P. Pillai, J. Nguyen, and B. Balasingam}}

\maketitle

\begin{abstract}
This paper is the third part of a series of papers about empirical approaches to open circuit voltage (OCV) modeling of lithium-ion batteries. 
The first part of the series \cite{SlowOCVp1} proposed models to quantify various sources of uncertainties in the OCV models; and, the second part of the series \cite{SlowOCVp2} presented systematic data collection approaches to compute the uncertainties in the OCV-SOC models. 
This paper uses data collected from 28 OCV characterization experiments, performed according to the data collection plan presented in \cite{SlowOCVp2}, to compute and analyze the following three different OCV uncertainty metrics: cell-to-cell variations, cycle-rate error, and curve fitting error.
From the computed metrics, it was observed that a lower C-Rate showed smaller errors in the OCV-SOC model and vice versa. 
The results reported in this paper establish a relationship between the C-Rate and the uncertainty of the OCV-SOC model. 
This research can be thus useful to battery researchers for quantifying the tradeoff between the time taken to complete the OCV characterization test and the corresponding uncertainty in the OCV-SOC modeling.
Further, quantified uncertainty model parameters can be used to accurately characterize the uncertainty in various battery management functionalities, such as state of charge and state of health estimation.
\end{abstract}



\section{Introduction}

Rechargeable Li-ion batteries have been adopted since 1991 in wide-ranging portable applications that require high energy capacity \cite{korthauer2018lithium}. 
The worldwide transition mandate from gasoline to electric vehicles has resulted in an elevated interest in Li-ion batteries \cite{haddadian2015accelerating}. 
In the automotive field, Li-ion batteries are being employed in hybrid electric vehicles (HEVs), battery electric vehicles (BEVs) and plug-in hybrid vehicles (PHEVs).
Several battery cells are arranged in a series, parallel or combined serial-parallel configuration to meet the voltage/power requirements. 
For monitoring of these packs, electric vehicles are equipped with a battery management system (BMS) \cite{wang2020comprehensive}.
The major functionality of a BMS is therefore monitoring the state of the battery and regulating its operation based on vital estimates depicting battery behavior such as the state of charge (SOC), battery capacity, and battery internal impedance.

\begin{figure}[h!]
\begin{center}
{\includegraphics[width=0.9\columnwidth]{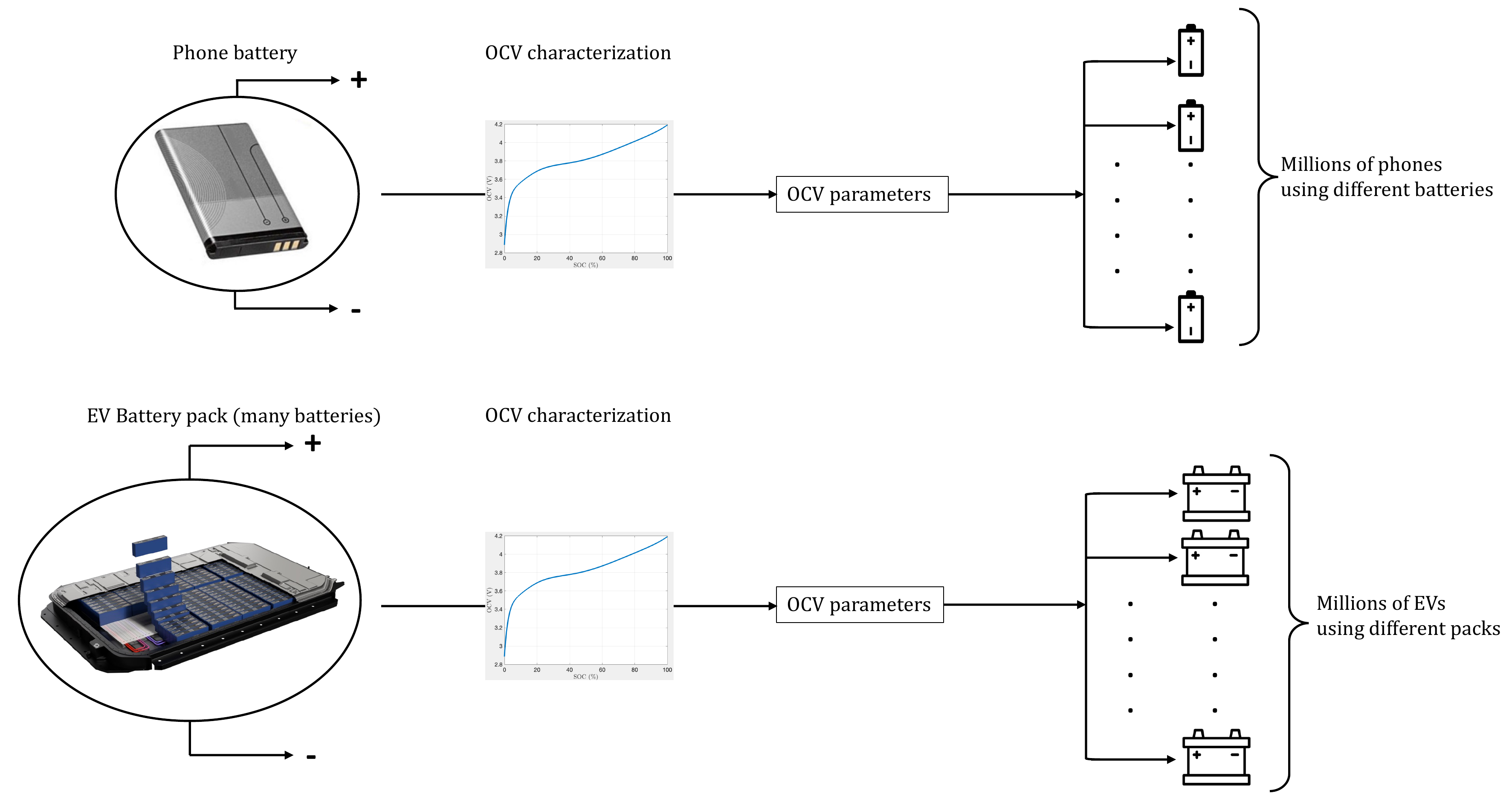}}
\caption{State of the art OCV Characterization.}
\label{fig:state-of-the-art-OCVchar}
\end{center}
\end{figure}

Typically, in most consumer electronic applications, the OCV-SOC characterization is obtained from a sample battery cell. It is then stored in the form of an OCV-SOC table \cite{sundaresan2022tabular} or a set of OCV-SOC parameters \cite{pillai2022open}. 
The same OCV-SOC curve (in the form of the parameters or the table) is used in millions of devices powered by batteries that are identical to the characterized one. 
In battery electric vehicle (BEV) applications, a sample battery pack (which is made out of thousands of cells) is characterized and the obtained parameters are used in BEVs that use identical battery packs. 
Figure \ref{fig:state-of-the-art-OCVchar} summarizes this state-of-the-art OCV-SOC characterization. 

The OCV-SOC curve is thought to be the same for identical batteries and battery packs \cite{chen2022novel}. 
However, various factors contribute to deviations of a certain battery's OCV characteristics from what is stored by the battery management system (which obtains the OCV parameters by characterizing another identical battery) \cite{rahimi2013battery}.

\begin{figure}[h!]
\begin{center}
{\includegraphics[width=0.6\columnwidth]{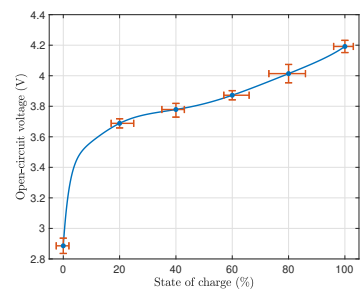}}
\caption{OCV Model Uncertainty.}
\label{fig:OCVerror}
\end{center}
\end{figure}

The blue line in Figure \ref{fig:OCVerror} shows the OCV-SOC curve stored by the battery management system; the OCV-SOC curve is obtained from a sample (identical) battery cell as illustrated in Figure \ref{fig:state-of-the-art-OCVchar}. 
The red lines indicate the possibility of discrepancies between the stored OCV-SOC curve and the realistic OCV-SOV characteristics of the battery being managed by the BMS. 
The uncertainty is possible in both (SOC and OCV) directions.
Assuming that the uncertainties can be considered to be a zero-mean error, the relationship between these two uncertainties is shown in \cite{SlowOCVp1} to be the following
\begin{align}
\sigma_\rs^2(s) =  \left (\frac{1}{f'(s)}\right)^2 \sigma_\rE^2 \label{eq:SOCerrorVar}
\end{align}
where
$f(s)$ denotes the OCV-SOC curve, 
$\sigma_\rs$ denotes the standard deviation (s.d.) of the uncertainty in the SOC direction, and 
$\sigma_\rE$ denotes the s.d. of the uncertainty in the OCV direction. 
Hence, the results and analysis presented in this paper will discuss the uncertainties only in the OCV direction.

The OCV-SOC characterization process is standardized differently for industry applications and is usually done on a sample battery. The OCV-SOC characterization is then stored as a table for use in multiple battery cells or packs.  
Usually, the characterization techniques are divided into two main categories: Galvanostatic Intermittent Titration Technique (GITT) and the low-rate cycling method \cite{wu2021comprehensive}.
In the GITT method, the sample battery is discharged in steps of 10\% until the entire SOC range is spanned.
At each step, a 10-second charge/discharge pulse is applied for the estimation of other battery parameters.
In the low-rate cycling method, a full sample battery is discharged until it becomes empty and it is then charged back again until the battery becomes full. The voltage and current data during the low-rate cycling method are recorded to employ curve-fitting techniques for OCV characterization. However, unlike the GITT technique, the low-rate cycling method is not standardized and the literature does not specify a standard current rate for performing this test.

The uncertainty in the stored OCV-SOC curve along the OCV direction can be due to many factors that arise during the characterization process and general battery behavior \cite{saha2019accurate}. The errors due to the characterization process can be listed as cell-to-cell variation, cycle rate effect, and curve-fitting error.
Other factors of error due to general characteristics of a battery with age and use are temperature variation and aging drift.
Currently, available BMS solutions fail to recognize the uncertainty in the OCV-SOC model and are predominantly focused on algorithms that reduce the estimation error on other battery parameters considering that the OCV-SOC model is perfectly known. This paper emphasizes that the error in the estimation of required battery parameters, such as the SOC, is affected by the uncertainties in the OCV-SOC relationship and these uncertainties are quantitatively analyzed for data recorded from Li-ion batteries in a laboratory setting.

The remainder of this paper is organized as follows: 
Section \ref{sec:data-collection} presents the details of the (identical) batteries used for the analysis and the details of the scientific grade high-precision data collection system.
Section \ref{sec:ocvsocdata} shows the details of the OCV-SOC characterization and 
Section \ref{sec:metrics} clearly defines the uncertainty metrics to be computed based on the collected data. 
Section \ref{sec:results} presents the three uncertainty measures, defined in Section \ref{sec:metrics}.
Section \ref{sec:data-analysis} presents a preliminary analysis of the collected data in terms of the computed battery capacity, resistance and the hysteresis effects of the identical cells studied in this paper.  
Finally, the paper is concluded in Section \ref{sec:conclusions}.

\section{Data Collection}
\label{sec:data-collection}

This section describes the data collection setup and the batteries used to collect the low-rate OCV test data in a laboratory setting. 

\subsection{Data Collection System}
\label{sec:Arbin}

\begin{figure}[h!]
\begin{center}
\subfloat[][Equipment.\label{fig:arbin}]
{\includegraphics[width=0.8\columnwidth]{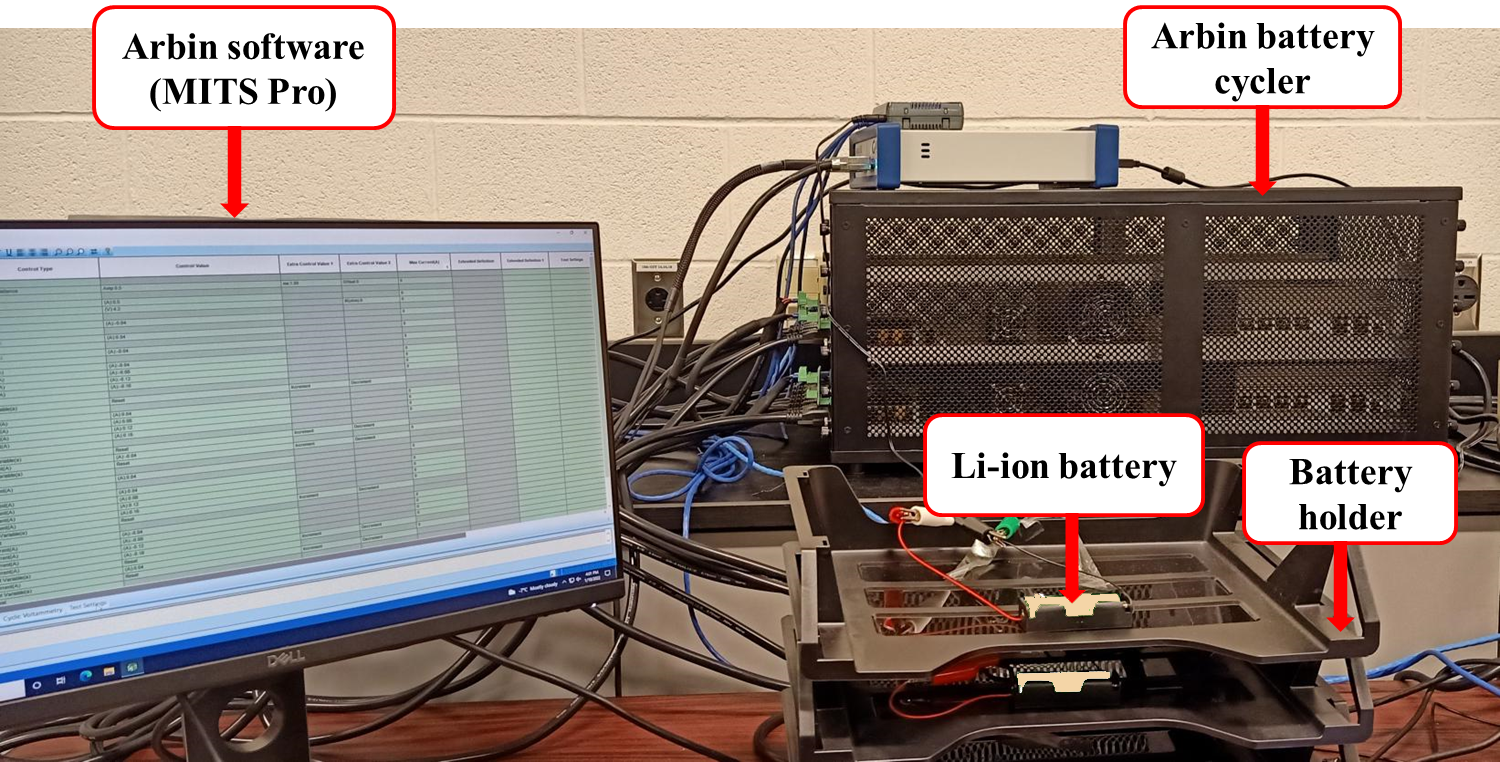}}\\
\subfloat[][Connection diagram.\label{fig:arbinblockdia}]
{\includegraphics[width=0.8\columnwidth]{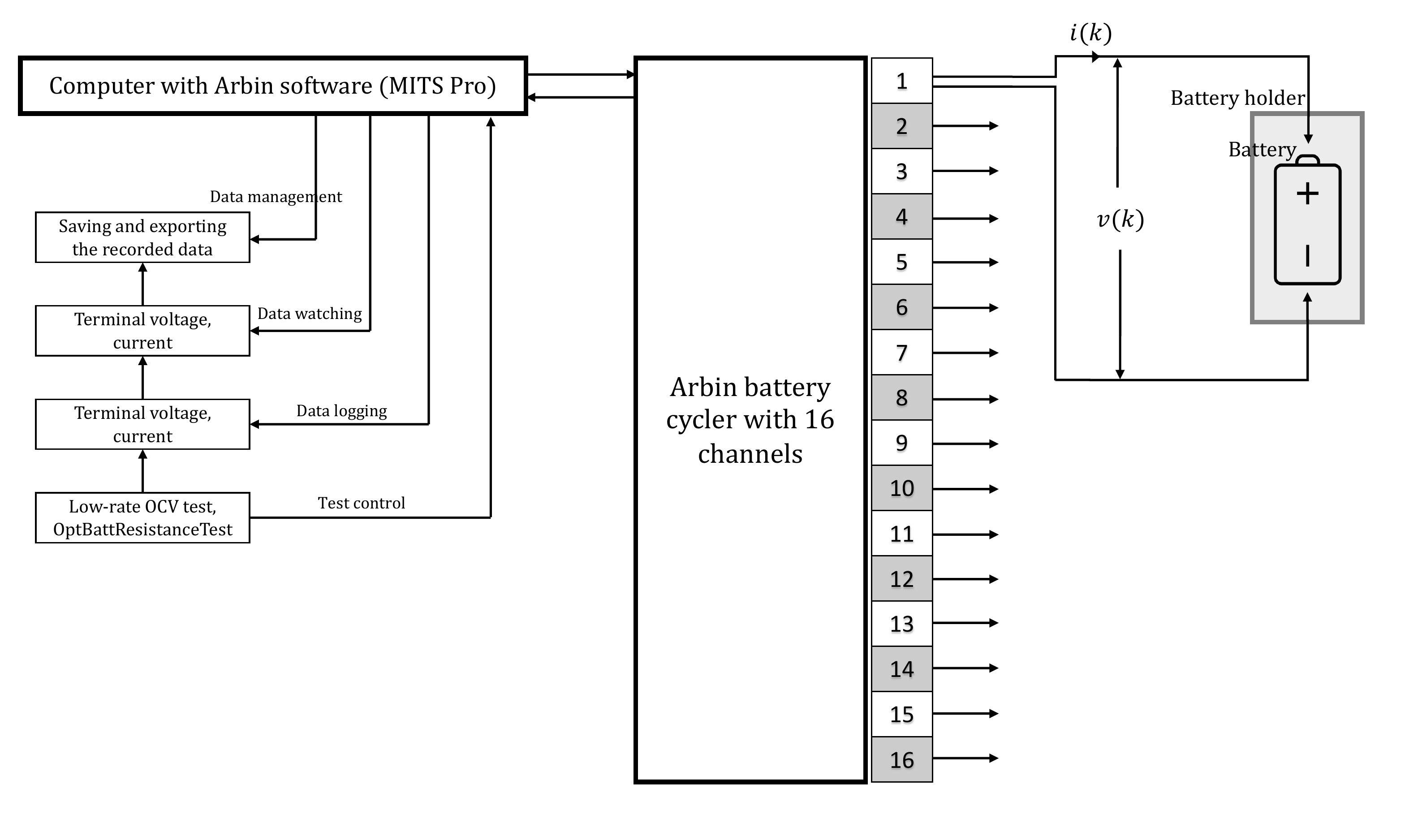}}
\caption{Arbin battery cycler.\label{fig:battsetup}}
\end{center}
\end{figure}

The voltage and current data from the battery were collected using a scientific-grade battery cycler made by Arbin Inc. The MITS Pro is Arbin's comprehensive battery testing software that allows programming each channel for a specific data collection plan. 
Figure \ref{fig:battsetup}\subref{fig:arbin} shows a picture of the Arbin laboratory battery testing (LBT) system which can be used to simultaneously cycle 16 batteries at a given time. 
Figure \ref{fig:battsetup}\subref{fig:arbinblockdia} shows a block diagram of the data collection system. 

\subsection{Batteries}

\begin{figure}[h!]
\begin{center}
{\includegraphics[width= 0.55\columnwidth]{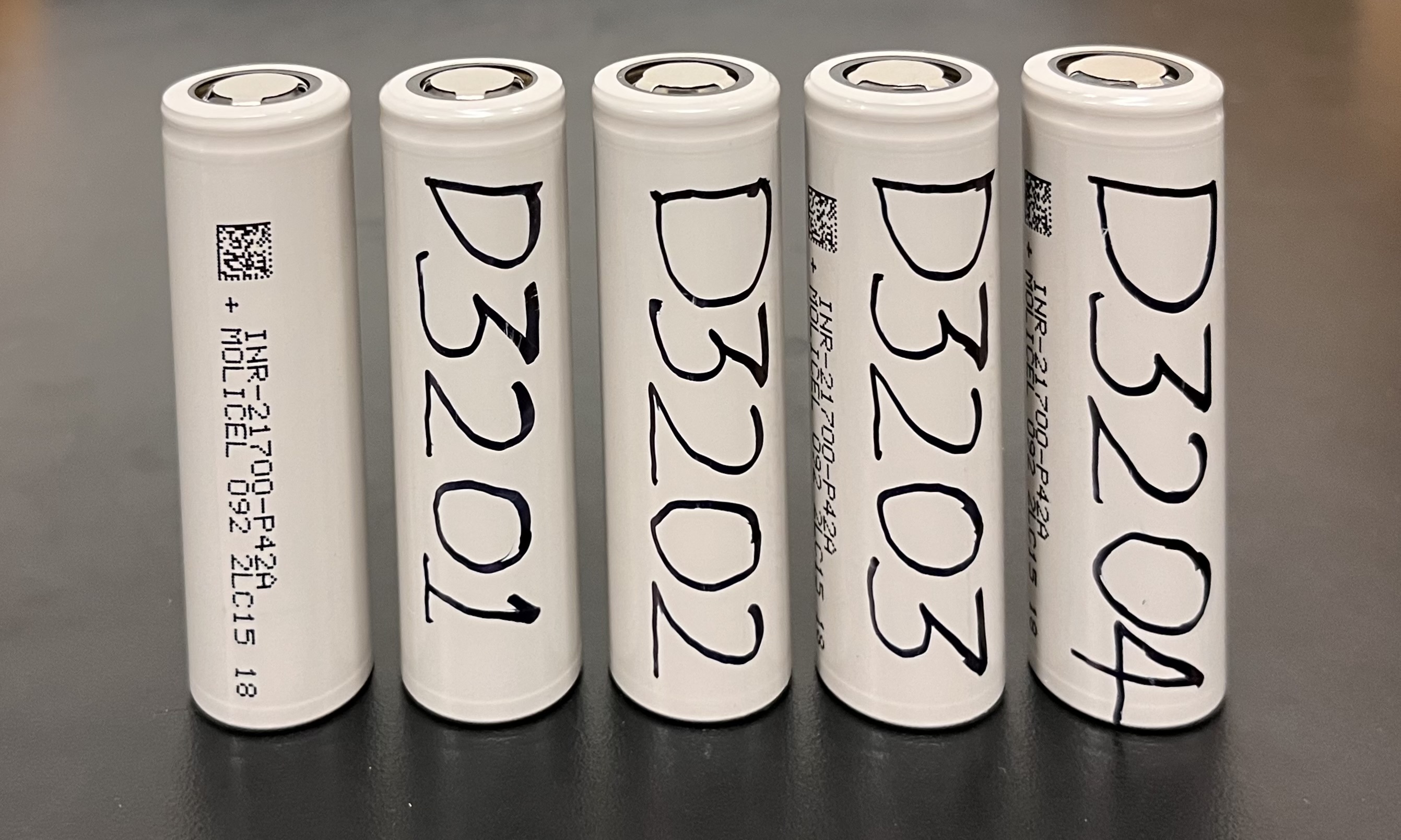}} \hspace{0.5cm}
\caption{Molicel INR-21700-P42A batteries.}
\label{fig:molicel}
\end{center}
\end{figure}

The data for the demonstration was based on Molicel INR-21700-P42A battery cells \cite{LImolicel}.
In total, 16 identical cells of the above serial number were used in the present study; these cells are labelled D3201, D3202, \ldots, and D3216 using a permanent marker. 
Figure \ref{fig:molicel} shows a picture of five cells used in the experiment. 
The nominal voltage of each battery cell is $3.6$V and their typical capacity and internal resistance are $4$ Ah and $16$ m$\Omega$, respectively. 

Table \ref{table:data-collection-plan} summarizes the entire data collection plan carried out in this work. 
In summary, four batteries were selected and the OCV characterization data was collected from them according to the data collection plan described by the routine `Low-Rate-OCV-Test'. 
The details and rationale of the low-rate OCV characterization data collection routine can be found in \cite{SlowOCVp2}; 
in short, the details of 
\begin{itemize}
\item[(a)] charging the battery before the experiment: the battery was first charged using a CC-CV approach with $C/N$ as the shutdown current.
\item[(b)] Then, the low-rate OCV test was carried out using a $C/N$ current. 
\item[(c)] Finally, a current pulse series was applied to estimate the resistance. 
\end{itemize}
are presented in the previous part of this paper \cite{SlowOCVp2}.
The data collection was repeated for seven different current rates (C/2 to C/128) at room temperature as shown in Table \ref{table:data-collection-plan}. 
Performance evaluation at different temperatures is deferred for future work. 

\begin{table}[h!]
\caption{{List of complete data collection}\label{table:data-collection-plan}}
\begin{center}
\begin{tabular}{|c|c|c|}
\hline
\textbf{Test} &  \textbf{Battery numbers} \\ \hline
{Low-Rate-OCV-Test} (C/2,Room) 	&	D3209,  D3210, D3211, D3212        \\ \hline
{Low-Rate-OCV-Test} (C/4,Room) 	&	D3205,  D3206, D3207, D3208        \\ \hline
{Low-Rate-OCV-Test} (C/8,Room) 	&	D3201,  D3202, D3203, D3204        \\ \hline
{Low-Rate-OCV-Test} (C/16,Room) 	&	D3201,  D3202, D3203, D3204        \\ \hline
{Low-Rate-OCV-Test} (C/32,Room)  	&	D3205,  D3206, D3207, D3208       \\ \hline
{Low-Rate-OCV-Test} (C/64,Room)  	&	D3209,  D3210, D3211, D3212        \\ \hline
{Low-Rate-OCV-Test} (C/128,Room)  	&	D3213,  D3214, D3215, D3216       \\ \hline
\end{tabular}
\end{center}
\end{table}

\subsection{Voltage-Current Data}

\begin{figure}[h!]
\begin{center}
\subfloat[][Terminal voltage during the C/2 data collection.]
{\includegraphics[width=0.9\columnwidth]{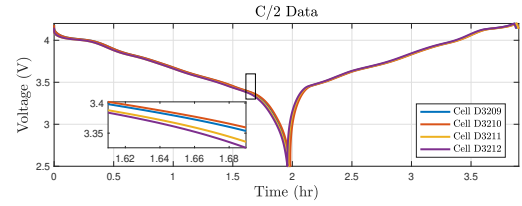}} \\
\subfloat[][Current during the C/2 data collection.]
{\includegraphics[width=0.9\columnwidth]{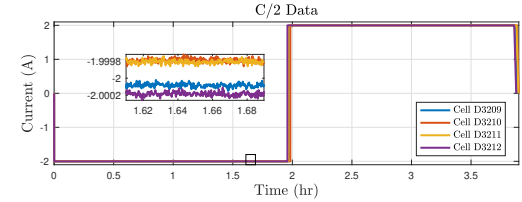}} \\
\subfloat[][Terminal voltage during the C/128 data collection.]
{\includegraphics[width=0.9\columnwidth]{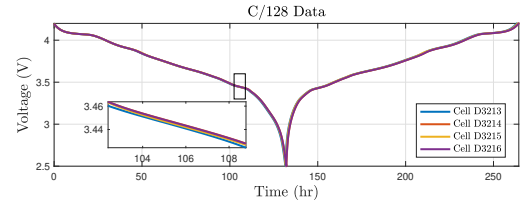}} \\
\subfloat[][Current during the C/128 data collection.]
{\includegraphics[width=0.9\columnwidth]{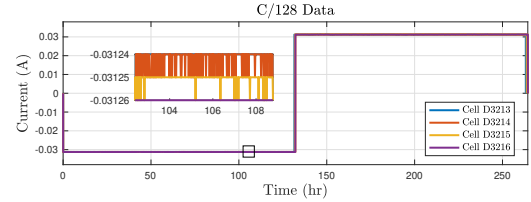}}
\caption{Low-rate OCV characterization data collected at the C/2 and C/128 rate.}
\label{fig:VIdata}
\end{center}
\end{figure}

Figure \ref{fig:VIdata} shows the voltage and current data for two different experiments (at C/2 rate and C/128 rate) listed in Table \ref{table:data-collection-plan}. 
Figure \ref{fig:VIdata}(a) and Figure \ref{fig:VIdata}(b) show the voltage and current data, respectively, collected during the low-rate OCV test at C/2 rate (i.e, using the data collection algorithm {Low-Rate-OCV-Test}(C/2,Room)). 
Similarly, Figure \ref{fig:VIdata}(c) and Figure \ref{fig:VIdata}(d) show the voltage and current data, respectively, collected using the data collection algorithm {Low-Rate-OCV-Test} (C/128, Room). 
A zoomed version of the curve shows the variations among the different C-Rates. It can be noticed from the zoomed portion that while the data collected at C/128 (from four different battery cells) overlays without many visible deviations, the data collected at the C/2 rate shows visible deviations. 
The goal of this series of papers is to quantify the effect of such differences on the ultimate performance of the BMS, particularly in SOC estimation.

\section{{Pseudo-OCV Modeling}}
\label{sec:ocvsocdata}

\begin{figure}[h!]
\begin{center}
{\includegraphics[width=0.9\columnwidth]{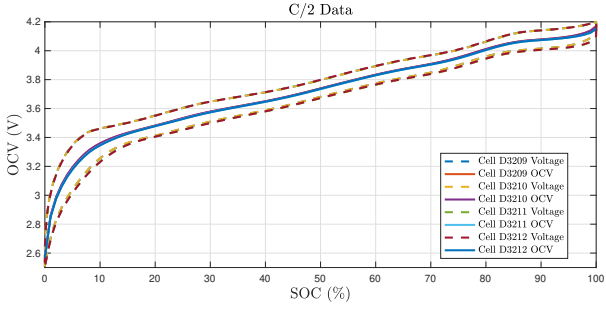}} \\
\caption{The OCV-SOC curves obtained through the combined+3 model \cite{pattipati2014open}. 
}
\label{fig:OCVplots}
\end{center}
\end{figure}

Given the voltage and current data (similar to the sample ones shown in Figure \ref{fig:VIdata}), the OCV-SOC curve can be obtained by averaging the charging and discharging terminal voltages.
Usually, the cell voltage is lower than the OCV during discharging and it is higher during charging. 
The OCV obtained by averaging the charging and discharging terminal voltages is expected to be the closest possible approximation of the true OCV-SOC curve of a battery. Figure \ref{fig:OCVplots} shows the OCV-SOC plot corresponding to the OCV experiment at $C/2$ rate for all four batteries. 

\begin{figure}[h!]
\begin{center}
{\includegraphics[width=0.9\columnwidth]{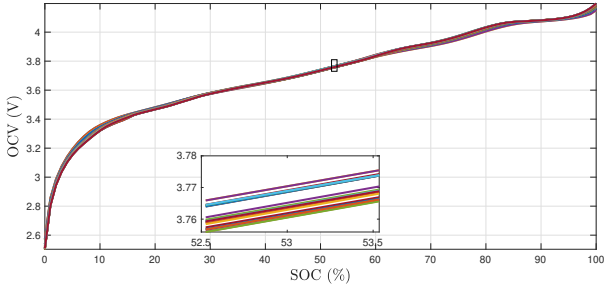}}
\caption{All 28 OCV-SOC plots.}
\label{fig:OCVmodelsAll}
\end{center}
\end{figure}

Let us take a closer look at the data and the OCV-SOC curve presented in Figure \ref{fig:OCVplots}. The cell voltage is shown using dashed lines and the averaged OCV is shown using solid lines. 
Figure \ref{fig:OCVmodelsAll} shows all the 28 OCV curves corresponding to the 28 experiments (7 C-Rates and 4 cells). 
A zoomed version of the curve shows the variations among these curves.
Subsequent sections of this paper will introduce formal approaches to compare the OCV curves obtained through experiments at different C-Rates. 

For the purpose of evaluating the OCV characterization, specific SOC values at equal intervals (also referred to as the SOC-Grid) are selected as below: 
\begin{align}
\bar \bs = 
\begin{bmatrix}
\bar s_1 & \bar s_2 & \ldots & \bar s_k
\label{eq:SOCgrid}
\end{bmatrix}^T
\end{align}
where $\bar s_2 - \bar s_1 = \bar s_3 - \bar s_2= \ldots \bar s_k - \bar s_{k-1}.$
Figure \ref{fig:OCVerror} shows a sample SOC-Grid for $k=6$.
For each SOC value in \eqref{eq:SOCgrid}, the corresponding OCV is then computed using linear interpolation based on the OCV characterization data depicted in Table \ref{table:OCVtable}.

\begin{table}[h!]
\caption{The OCV-SOC Table}
\label{table:OCVtable}
\begin{center}
\begin{tabular}{|C{1 in}|C{1 in}|}
\hline 
SOC & OCV \\ \hline
\end{tabular}\\
\begin{tabular}{|L{1 in}|L{1 in}|}
$s_1=0$ & $v_1 = {\rm OCV_{min}}$ \\ \hline
$s_2$ & $v_2$ \\ \hline
\vdots & \vdots   \\ \hline
$s_n=1$ & $v_n = {\rm OCV_{max}}$ \\ \hline
\end{tabular}
\end{center}
\end{table}

\section{Computation of OCV Uncertainty Metrics}
\label{sec:metrics}

In this section, the procedure to compute the three uncertainty metrics is described. Table \ref{table:OCVtable} represents the data obtained as a result of the OCV-SOC data characterization (see \cite{SlowOCVp2} for details) from one battery cell at a certain C-Rate. 
In this paper, such data obtained from several battery cells at several C-Rates will be formally evaluated to understand the following three OCV uncertainty metrics defined in \cite{SlowOCVp2}:
cell-to-cell variations, 
cycle rate error, and curve fitting error.

\subsection{Cell-to-Cell Variation Metrics}
\label{sec:c2c-error}

First, the OCV characterization approach presented in \cite{SlowOCVp2} needs to be followed to compute the SOC and OCV data (similar to the one indicated in Table \ref{table:OCVtable}). This procedure is then repeated for all the cells and Table \ref{table:C2CvarOCV} shows the resulting validation data for $q$ cells. 
The cell-to-cell variation metrics are then computed as follows 
\begin{align}
\mu_{\rm c2c} (\bar s_l ) 
&=  \frac{1 }{ \binom q2 }  
\left(  \sum \limits_{i=1, \, j =i+1, \, i \neq j }^{i=q-1, \, j =q} \left( v_l(i) -  v_l(j)  \right)\right) \label{eq:mu-c2c} \\
\sigma_{\rm c2c} (\bar s_l ) 
& =  \sqrt{ \frac{1}{ \binom q2 } \left(  \sum \limits_{i=1, \, j =i+1; \, i \neq j }^{i=q-1, \, j = q} \left(  v_l(i) -  v_l(j)  \right)^2 \right)}
\label{eq:std-c2c}
\end{align}
where $\binom q 2$ denotes the possible number of combinations and the grid number $l$ ranges from 1 to $k$.
 
\begin{remark}
Let us assume that four batteries were tested for their cell-to-cell variance, i.e., $q=4$.
Consider, battery `1' is chosen as the sample battery. Using this sample battery, the error due to cell-to-cell variance is computed for combinations ($1, 2$), ($1, 3$), and ($1, 4$). 
Similarly, if battery `2' is chosen as the sample battery, then the cell-to-cell variance is computed for ($1, 2$), ($2, 3$), and ($2, 4$) and when battery `3' is chosen as the sample battery, the cell-to-cell variance is computed for ($1, 3$), ($2, 3$), and ($3, 4$).
Thus, if any of the four batteries are chosen as the sample battery for OCV characterization, one can compute the cell-to-cell variance for all 6 non-repeating combinations, i.e., ($1, 2$), ($1, 3$), ($1, 4$), ($2, 3$), ($2, 4$), and ($3, 4$).
Thus, in \eqref{eq:mu-c2c}, the cell-to-cell variation is computed for all combinations of $i$ and $j$ i.e., the uncertainty due to cell-to-cell variation when any battery from the given set of $q$ batteries are chosen as the sample battery. 
\end{remark}

\begin{table}[h!]
\caption{Evaluation Data for Cell-to-Cell OCV Variation}
\label{table:C2CvarOCV}
\begin{center}
\begin{tabular}{|c|c|c|c|c|c|}
\hline 
${\rm SOC}$   & ${\rm OCV}$ (cell-1) & $\ldots$ & ${\rm OCV}$(cell-q) \\ \hline
$\bar s_1$ &   $  v_1(1) $ &  $\ldots$ &  $  v_1(q)$\\ \hline
$\bar s_2$ &  $   v_2 (1)$ &  $\ldots$ &  $ v_2(q)$\\ \hline
\vdots  & \vdots & \vdots & \vdots \\ \hline
$\bar s_k$ &  $  v_k(1) $ &  $\ldots$ &  $ v_k(q)$\\ \hline
\end{tabular}
\end{center}
\end{table}

So far, all computed errors due to cell-to-cell variation, such as the mean $\mu_{\rm c2c} (\bar s_l )$, is calculated for a particular SOC, $\bar s_l$. Thus, the average of the mean and standard deviation for a given C-Rate is calculated as follows:
\begin{align}
\mu_{\rm c2c} =  \frac{1}{k} \sum \limits_{l=1 }^{k} \mu_{\rm c2c} (\bar s_l ); \quad \quad \sigma_{\rm c2c} =  \frac{1}{k} \sum \limits_{l=1 }^{k} \sigma_{\rm c2c} (\bar s_l ) \label{eq:C2Cavg}
\end{align}

It is important to note that the validation data shown in Table \ref{table:C2CvarOCV}, and the validation metrics computed in \eqref{eq:mu-c2c}, \eqref{eq:std-c2c}, and \eqref{eq:C2Cavg} are for a particular C-Rate. 
The procedure must be repeated for each C-Rate to see if the cell-to-cell variation metrics change with C-Rate.

\subsection{C-Rate Error Metrics }
\label{sec:C-Rate-error}

Computation of C-Rate error starts with the data format presented in Table \ref{table:C2CvarOCV} (which is shown for a particular C-Rate).
First, to minimize the cell-to-cell effects, the data from all the $q$ cells are averaged as follows
\begin{align}
\bar v_l (\text{C-Rate})= \frac{ \sum_{i=1}^{q} v_l(i) }{q} \quad \forall  \quad l = 1, \ldots, k
\label{eq:refOCV}
\end{align}
Table \ref{table:C-RateVar} shows the data that need to be prepared to compute the uncertainties with respect to the C-Rate.  
The second column in Table \ref{table:C-RateVar} shows the averaged OCV for the reference C-Rate, which is relatively the lowest. 
Here, the reference C-Rate (denoted as `cr-ref') is the lowest C-Rate experiment, i.e., `cr-ref' = C/128. The averaged OCV data needs to be computed for all the C-Rate as shown in the remaining columns of Table \ref{table:C-RateVar}. 
\begin{table}[h!]
\caption{Evaluation Data for C-Rate Variation}
\label{table:C-RateVar}
\begin{center}
\begin{tabular}{|c|c|c|c|c|c|}
\hline 
${\rm SOC}$   & ${\rm OCV}$ (ref)  & ${\rm OCV}$ (C-Rate-1) & $\ldots$ & ${\rm OCV}$(C-Rate-r) \\ \hline
$\bar s_1$ &   $ \bar v_1 $(cr-ref) &   $ \bar v_1(1) $ &  $\ldots$ &  $\bar v_1(r)$\\ \hline
$\bar s_2$ &  $ \bar v_2 $(cr-ref) &  $ \bar v_2 (1)$ &  $\ldots$ &  $\bar v_2(r)$\\ \hline
\vdots  & \vdots & \vdots & \vdots & \vdots \\ \hline
$\bar s_k$ &  $\bar v_k $(cr-ref) &  $\bar v_k(1)  $ &  $\ldots$ &  $\bar v_k(r)$\\ \hline
\end{tabular}
\end{center}
\end{table}

Considering the data collected from $q$ cells shown in Table \ref{table:C-RateVar}, the C-Rate variance of OCV is computed as follows
 
\begin{align}
\mu_{\rm cr} (\bar s_l ) &=  \frac{1}{r} \left( \sum \limits_{i=1 }^{r} \left( \bar v_l(i) - \bar v_l(\text{cr-ref})  \right) \right) 
\, \forall  \quad l = 1, \ldots, k
 \label{eq:mean-cr} \\ 
\sigma_{\rm cr} (\bar s_l ) &=  \sqrt{ \frac{ \sum \limits_{i =1 }^{r} \left( \bar v_l(i) - \bar v_l(\text{cr-ref})  \right)^2 }{ r-1} }
\quad \forall  \quad l = 1, \ldots, k
\label{eq:std-cr}
\end{align}

All the previously computed errors due to C-Rate variation, such as the mean $\mu_{\rm cr} (\bar s_l )$, are calculated for a particular SOC, $\bar s_l$. Thus, the average of the mean and standard deviation due to a given C-Rate is calculated as follows:
\begin{align}
\mu_{\rm cr} =  \frac{1}{k} \sum \limits_{l=1}^{k} \mu_{\rm cr} (\bar s_l ); \quad \quad \sigma_{\rm cr} =  \frac{1}{k} \sum \limits_{l=1}^{k} \sigma_{\rm cr} (\bar s_l ) \label{eq:CRateavg}
\end{align}

\subsection{Curve Fitting Error Metrics}

The evaluation of curve-fitting models and approaches to OCV modeling has been extensively studied in the past. 
In \cite{pattipati2014open,pillai2022open}, several OCV models (which fall under the category of linear, non-linear, hybrid, and tabular models) were evaluated based on several metrics. 
In this paper, the curve fitting error is evaluated based on the mean and standard deviation of the error, in line with the other uncertainties defined in Section \ref{sec:c2c-error} and Section \ref{sec:C-Rate-error}.

Table \ref{table:CFerror} shows the evaluation data that must be prepared to compute the curve-fitting error metrics. 
Here, the second column denotes the reference OCV, which is the average of OCV values across multiple cells, as defined in \eqref{eq:refOCV}. 
The remaining columns in Table \ref{table:CFerror}, from the third column to the last, are corresponding OCV values computed from the OCV function as follows:
\begin{align}
&\tilde v_l(j) =  \frac{1}{q} \sum\limits_{i=1}^{q}  f_{\rm em} (\bar s_l, \bk_\ro^i, i, j), \label{eq:curvefittingem} \\ \nonumber 
&\forall    \quad l = 1, \ldots, k \quad  \text{and}  \quad \forall   \quad j = 1, \ldots, s 
\end{align}
where
$f_{\rm em} (\bar s_l, \bk_\ro^i, i, j)$ denotes the OCV computed using the empirical OCV function for the curve fitting approach $j$. $\bk_\ro^i$ denotes the OCV parameters of the chosen battery $i$, and is generally written as
$\bk_\ro^i = \{ k_1^i, k_2^i, k_3^i, \dots, k_n^i\}$.
$s$ denotes the number of curve-fitting models to be evaluated.

\begin{table}[h!]
\caption{Evaluation Data for Curve Fitting Error}
\label{table:CFerror}
\begin{center}
\begin{tabular}{|c|c|c|c|c|c|}
\hline 
${\rm SOC}$   & ${\rm OCV}$ (ref)  & ${\rm OCV}$ (CF-1) & $\ldots$ & ${\rm OCV}$(CF-s) \\ \hline
$\bar s_1$ &   $ \bar v_1$(cr-ref) &   $ \tilde v_1(1) $ &  $\ldots$ &  $\tilde v_1(s)$\\ \hline
$\bar s_2$ &  $ \bar v_2 $(cr-ref) &  $ \tilde v_2 (1)$ &  $\ldots$ &  $\tilde v_2(s)$\\ \hline
\vdots  & \vdots & \vdots & \vdots & \vdots \\ \hline
$\bar s_k$ &  $\bar v_k $(cr-ref) &  $\tilde v_k(1)  $ &  $\ldots$ &  $\tilde v_k(s)$\\ \hline
\end{tabular}
\end{center}
\end{table}

Considering the data collected from $q$ cells shown in Table \ref{table:C-RateVar}, the curve-fitting metrics for OCV uncertainty are computed as follows 
\begin{align}
\mu_{\rm cf} (\bar s_l ) & =   \tilde v_l(j) -  \bar v_l({\rm cr})  
\quad \forall  \quad l = 1, \ldots, k \label{eq:mean-cf} \\ 
\sigma_{\rm cf} (\bar s_l ) & = \left( \tilde v_l(j) -  \bar  v_l({\rm cr})  \right)^2 
\quad \forall  \quad l = 1, \ldots, k
\label{eq:std-cf}
\end{align}
The average error for a given C-Rate is calculated as follows:
\begin{align}
\mu_{\rm cf} =  \frac{1}{k} \sum \limits_{l=1 }^{k} \mu_{\rm cf} (\bar s_l ); \quad \quad \sigma_{\rm cf} =  \frac{1}{k} \sum \limits_{l=1 }^{k} \sigma_{\rm cf} (\bar s_l ) \label{eq:CFavg}
\end{align}
The curve fitting error metrics are computed for a particular C-Rate (referred to as `cr') in \eqref{eq:mean-cf}, \eqref{eq:std-cf} and \eqref{eq:CFavg}. The procedure is then repeated for all C-Rates.

\section{Analysis of OCV Uncertainty Measures}
\label{sec:results}

In this section, the three OCV uncertainty metrics: cell-to-cell variation, cycle rate effect, and curve-fitting error defined in Section \ref{sec:metrics} are calculated for the OCV-SOC data from Section \ref{sec:ocvsocdata}.
In all the analyses presented below, a value of  $k=100$ is chosen for the SOC-Grid in \eqref{eq:SOCgrid}.

\subsection{Test for zero-mean}
In general, it is assumed that the OCV uncertainty is a zero-mean Gaussian variable, i.e., $\tilde E \sim \cN (0, \sigma_\rE^2 )$. 
First, we will check if the mean of the metrics is zero. 
This assumption can be verified by testing the means of the three computed metrics: 
cell-to-cell variations $\mu_{\rm c2c} $ from \eqref{eq:C2Cavg}, cycle-rate error $\mu_{\rm cr}$ from \eqref{eq:CRateavg}, and curve fitting error $\mu_{\rm cf}$ from \eqref{eq:CFavg}.
%
%
%
%
Let us define the standard mean of an uncertainty metric as 
\begin{align}
\hat \mu_{\rm c2c} = \frac{ \mu_{\rm c2c}}{ \sigma_{\rm c2c}} \label{eq:stdmean}
\end{align}

\begin{figure}[h!]
\begin{center}
{\includegraphics[width=0.8\columnwidth]{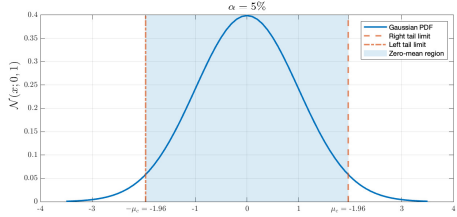}}
\caption{Standard Gaussian test for zero-mean.}
\label{fig:siglevel}
\end{center}
\end{figure}

To confirm the hypothesis that the standard mean above is zero, the following inequality must become true for a particular significance level,
\begin{align}
-\mu_c < \hat \mu_{\rm c2c} &< \mu_c \label{eq:sign}
\end{align}
For a significance level of $\alpha = 5\%$, the critical mean is $\mu_c = 1.96$ and this value is obtained from the standard normal distribution table as shown in Figure \ref{fig:siglevel}. Using similar steps, the zero-mean hypothesis tests for the other OCV uncertainty metrics, C-Rate and curve fitting can also be calculated using the standard mean values similar to \eqref{eq:stdmean}.

Finally, each cell is independent of each other. Hence, by considering a large number of SOC-grid points $k$, and based on the central limit theorem \cite{bar2001estimation}, the error can be assumed to be Gaussian.

\subsection{Cell-to-Cell Variations}

\begin{figure}[h!]
\begin{center}
\subfloat[][$\mu_{\rm c2c} (\bar s_l ) $ comparison for C/2 and C/128 C-Rates.\label{fig:C2C_16}]
{\includegraphics[width=0.8\columnwidth]{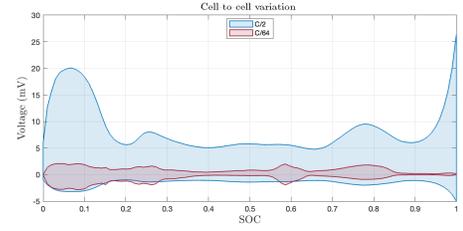}}\\
\subfloat[][$\mu_{\rm c2c} (\bar s_l ) $ comparison for C/4 and C/32 C-Rates.\label{fig:C2C_25}]
{\includegraphics[width=0.8\columnwidth]{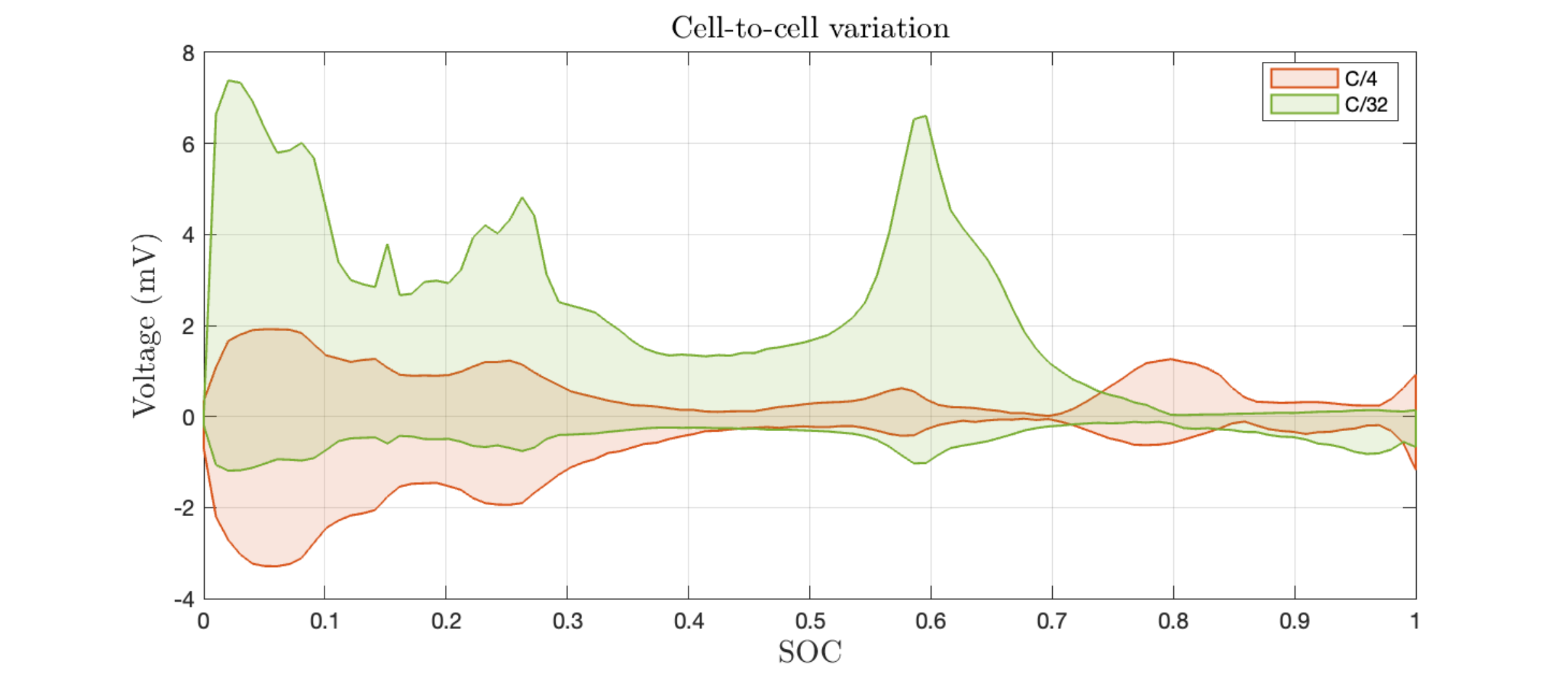}}\\
\subfloat[][$\mu_{\rm c2c} (\bar s_l ) $ comparison for C/8 and C/16 C-Rates.\label{fig:C2C_34}]
{\includegraphics[width=0.8\columnwidth]{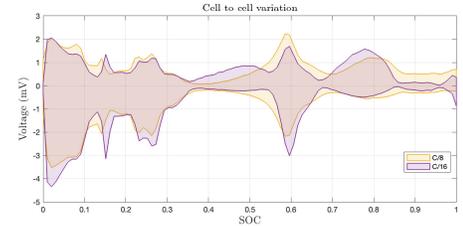}}
\caption{Cell-to-cell variations.\label{fig:C2Cresults}}
\end{center}
\end{figure}

From Table \ref{table:data-collection-plan}, the same four batteries, D3209-D3212 were cycled at C/2 and then at C/64. 
Their cell-to-cell variations for these two C-Rates were compared. In Figure \ref{fig:C2Cresults}\subref{fig:C2C_16}, the cell-to-cell variation is calculated for the OCV-SOC data from C/2 and C/64 rates as defined in \eqref{eq:mu-c2c}. The solid fill in this figure indicates the region spanned by ($\mu_{\rm c2c} + \sigma_{\rm c2c}$) and ($\mu_{\rm c2c} - \sigma_{\rm c2c}$). This filled region is denoted as the window of uncertainty and $2\sigma_{\rm c2c}$ serves as its width.

In Figure \ref{fig:C2Cresults}\subref{fig:C2C_25}, the cell-to-cell variation among C/4 and C/32 is compared using the data from battery cells D3205-D3208.
 The mean cell-to-cell variation is calculated using \eqref{eq:mu-c2c} and the standard deviation of the cell-to-cell variation is calculated using \eqref{eq:std-c2c} for both C-Rates.

Figure \ref{fig:C2Cresults}\subref{fig:C2C_34} shows the cell-to-cell variation between batteries D3201-D3204. The low-rate OCV test was conducted on the four batteries using a C/8 rate and then using a C/16 rate. The solid fill in this figure indicates the standard deviation of the cell-to-cell variation calculated using \eqref{eq:std-c2c}. 
An observation from the windows of uncertainty in Figure \ref{fig:C2Cresults} is that the width of uncertainty did not have any observable relationship with C-Rate. However, the width of uncertainty significantly increased at the C/2 rate.

\begin{table}[h!]
\begin{center}
\caption{Average error in mean and standard deviation of OCV uncertainty for seven C-Rates.\label{tabl:avgc2c}}
\begin{tabular}{|c|c|c|c|}
\hline
C-Rate      & $\mu_{\rm c2c}$ (mV) & $\sigma_{\rm c2c}$ (mV) & $\hat \mu_{\rm c2c}$ \\ \hline
C/2   & 3.4464                        &       5.0489        &       0.6826          \\ \hline
C/64  & 0.0965        &    0.9236       &    0.1045      \\ \hline \hline
C/4   & -0.1197        &     0.7705  &    -0.1553     \\ \hline
C/32  & 0.9224       &        1.4126              &   0.653    \\ \hline \hline
C/8   & -0.0749          &     0.9161    &   -0.0818      \\ \hline
C/16  & -0.1301        &     0.8799    &    -0.1479  \\ \hline \hline
C/128 & -0.2528       &     1.0313   &    -0.2452 \\ \hline
\end{tabular}
\end{center}
\end{table}

Table \ref{tabl:avgc2c} shows the mean cell-to-cell uncertainty for all the seven C-Rates. This table confirms that the width of the uncertainty significantly increased at a C/2 rate.
It can also be seen that the standard mean calculated for the OCV uncertainty due to cell-to-cell variation from \eqref{eq:stdmean} in Table \ref{tabl:avgc2c} conforms to the inequality for zero-mean significance. Hence, it can be confirmed with 95\% confidence that the error in OCV uncertainty due to cell-to-cell variation is zero-mean for all C-Rates.

\subsection{Cycle-Rate Error}

\begin{figure}[h!]
\begin{center}
{\includegraphics[width=0.8\columnwidth]{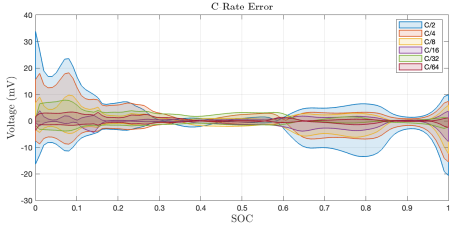}}
\caption{Cycle-rate error for six C-Rates with C/128 as the reference C-Rate (cr-ref).\label{fig:meanstd-cr}}
\end{center}
\end{figure}

In Figure \ref{fig:meanstd-cr}, the C-Rate uncertainties computed using \eqref{eq:mean-cr} and \eqref{eq:std-cr} for the six C-Rates are shown. The lowest C-Rate of C/128 is chosen as the reference curve, `cr-ref'. The uncertainty in OCV for the other six rates: C/2, C/4, C/8, C/16, C/32, and C/64 are calculated for a 100-point grid between SOC=0 and SOC=1. 
It can be observed that the uncertainty in OCV increases with increasing C-Rate. The highest error is observed for the C/2 rate, followed by C/4 and so on. The C/32 rate does not conform to the expected pattern and is considered an outlier in the current analysis.

The width of uncertainty, $2\sigma_{\rm c2c}$, for the window of uncertainty between ($\mu_{\rm c2c} + \sigma_{\rm c2c}$) and ($\mu_{\rm c2c} - \sigma_{\rm c2c}$) is not the same for all SOC. A higher width is seen at lower and higher SOC ranges. Further, in almost all comparisons of cycle-rate error, an increase in the width of uncertainty can also be seen for SOC values between 0.6 and 0.85.

Table \ref{tabl:avgcr} shows the mean cycle rate uncertainty for all the seven C-Rates. It can be noticed that the window of uncertainty consistently increased with the C-Rate. Further, the table also contains the values of the standard mean calculated by dividing the mean by the standard deviation. Based on the magnitude of the standard mean values and the inequality for significance in \eqref{eq:sign}, it can be confirmed with 95\% that the OCV uncertainty due to the C-Rate is also zero-mean Gaussian.

\begin{table}[h!]
\begin{center}
\caption{Average error in mean and standard deviation of OCV uncertainty for six C-Rates.\label{tabl:avgcr}}
\begin{tabular}{|c|c|c|c|}
\hline
C-Rate      & $\mu_{\rm cr}$ (mv) & $\sigma_{\rm cr}$ (mV) & $\hat \mu_{\rm cr}$  \\ \hline
C/2    & -0.045       &      5.7236          &    -0.0079            \\ \hline
C/4    & 0.3057      &          4.1244       &       0.0741        \\ \hline
C/8    & -0.1568        &       2.5097      &       -0.0625            \\ \hline
C/16  & -0.3159        &        1.2214      &       -0.2586         \\ \hline 
C/32  & 0.5652        &         1.9269      &         0.2933        \\ \hline 
C/64  & 0.1474         &      0.8661    &            0.1702          \\ \hline 
\end{tabular}
\end{center}
\end{table}

\subsection{Curve Fitting Error}

In this section, three curve-fitting approaches: Nernst, Combined and Combined+3 models \cite{pillai2022open} for OCV are evaluated with the reference OCV. The curve fitting error for a general case is given in \eqref{eq:curvefittingem}. This equation is adopted for the three models as follows: \\
\textbf{Nernst model}
\begin{align}
\tilde v_l(1) &=  \frac{1}{4} \sum\limits_{i=1}^{4}  f_{\rm em} (\bar s_l, \bk_\ro^i, i, 1) \\
&=  \frac{1}{4} \sum\limits_{i=1}^{4} \left( k_0^i + k_1^i \ln(\bar s_l) +  k_2^i \ln(1-\bar s_l)  \right)  \label{eq:Nernst}
\end{align}
\textbf{Combined model}
\begin{align}
\tilde v_l(2) &=  \frac{1}{4} \sum\limits_{i=1}^{4}  f_{\rm em} (\bar s_l, \bk_\ro^i, i, 2) \\
&= \frac{1}{4} \sum\limits_{i=1}^{4}  \left( k_0^i + \frac{k_1^i}{\bar s_l} + k_2^i \bar s_l  + k_3^i \ln(\bar s_l) + k_4^i\ln(1-\bar s_l) \right)  \label{eq:combined}
\end{align}
\textbf{Combined+3 model}
\begin{align}
\tilde v_l(3) &=  \frac{1}{4} \sum\limits_{i=1}^{4}  f_{\rm em} (\bar s_l, \bk_\ro^i, i, 3) \\
&=   \frac{1}{4} \sum\limits_{i=1}^{4}  \left(  k_0^i +  \frac{k_1^i}{\bar s_l} +   \frac{k_2^i}{\bar s_l^2} + \frac{k_3^i}{\bar s_l^3} + \frac{k_4^i}{\bar s_l^4} \right. \nonumber \\
& \quad \quad \quad + k_5^i\bar s_l +  k_6^i\ln(\bar s_l) + k_7^i \ln(1-\bar s_l) \biggl)  \label{eq:combined+3}
\end{align}

\begin{figure}[h!]
\begin{center}
\subfloat[][Curve fitting error for C/2 rate.\label{fig:c2CF}]
{\includegraphics[width=0.8\columnwidth]{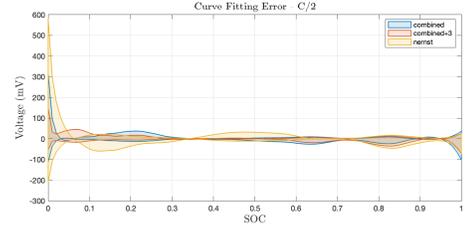}}\\
\subfloat[][Curve fitting error for C/128 rate.\label{fig:c128CF}]
{\includegraphics[width=0.8\columnwidth]{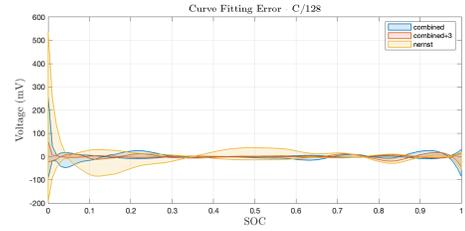}}
\caption{Curve-fitting error calculated for three empirical models: Nernst, combined, combined+3.\label{fig:meanstd-cf}}
\end{center}
\end{figure}

The mean and standard deviation are calculated for each C-Rate using \eqref{eq:mean-cf} and \eqref{eq:std-cf}, respectively, for all three models. Figure \ref{fig:meanstd-cf}\subref{fig:c2CF} and \ref{fig:meanstd-cf}\subref{fig:c128CF} shows the uncertainty in OCV for the C/2 and C/128 rates. It can be observed from these two figures that the width of the curve-fitting error for a particular empirical model appears to be the same across differing C-Rates.

Table \ref{tabl:avgcf} shows the mean uncertainty for all the seven C-Rates. Similar to previous metrics, the mean of the OCV uncertainty due to curve fitting approaches was also confirmed to be zero-mean with 95\% confidence.

\begin{table}[h!]
\begin{center}
\caption{Average error in mean and standard deviation of OCV uncertainty due to three curve-fitting approaches.\label{tabl:avgcf}}
\begin{tabular}{|c|c|l|l|l|} \hline
C-Rate   & Empirical model  & \multicolumn{1}{c|}{$\mu_{\rm cf} $ (mV)} & \multicolumn{1}{c|}{$\sigma_{\rm cf}$ (mV)} &  \multicolumn{1}{c|}{$\hat \mu_{\rm cf}$ } \\ \hline
\multirow{3}{*}{C/2}   &   Combined   &    0.296     & 13.0189   &  0.0227 \\ \cline{2-5} 
                      		& Combined+3   &      0.0515          &          10.5123  &   0.0049        \\ \cline{2-5} 
                      		&  Nernst &     0.7709                     &       23.7226     &   0.0325      \\ \hline \hline
\multirow{3}{*}{C/4}  &  Combined    &   1.1434                           &        11.1291           &  0.1027   \\ \cline{2-5} 
                      		& Combined+3   &          0.8712                 &      8.1158         &    0.1073   \\ \cline{2-5} 
                      		&  Nernst &   1.6366    &     23.9871   &   0.0682     \\ \hline \hline
\multirow{3}{*}{C/8}  &   Combined   &      0.0563         &11.0874   & 0.0051 \\ \cline{2-5} 
                      		& Combined+3   &     -0.2168                    &    6.3627   &  -0.0341\\ \cline{2-5} 
                      		&  Nernst&         0.5676                           &    23.8462           &   0.0238                     \\ \hline \hline
\multirow{3}{*}{C/16} & Combined      &    -0.3188         &        10.4843 &  -0.0304  \\ \cline{2-5} 
                      		& Combined+3   &    -0.6102                   &          5.5187         &         -0.1106          \\ \cline{2-5} 
                      		& Nernst&       0.1781                               &       24.6426        &    0.0072                 \\ \hline \hline
\multirow{3}{*}{C/32} &Combined      &     1.7389                  &       10.1947          &   0.1706                     \\ \cline{2-5} 
                      		& Combined+3   & 1.4385                                 &         6.1602      &    0.2335                   \\ \cline{2-5} 
                      		& Nernst&    2.244                     &               24.7088                    & 0.0908     \\ \hline \hline
\multirow{3}{*}{C/64} & Combined      &   0.7561           &    10.2592        &        0.0737                   \\ \cline{2-5} 
                      		& Combined+3   &     0.4633                    &         5.4075       &    0.0857                     \\ \cline{2-5} 
                      		& Nernst &     1.2671                                 &           24.6134           &     0.0515           \\ \hline \hline
\multirow{3}{*}{C/128} & Combined      &     0.3664        &    10.2656      &         0.0357                   \\ \cline{2-5} 
                      		 & Combined+3   &       0.0856                   &        5.3333      &   0.0161                     \\ \cline{2-5} 
                      		& Nernst&     0.8583                                   &      25.0316          &   0.0343                    \\ \hline
\end{tabular}
\end{center}
\end{table}

\section{Other Analysis}
\label{sec:data-analysis}
In this section, additional analysis on the battery capacity, internal resistance and hysteresis using the collected data is presented. Some theoretical background for the following subsections is provided in part-2 of the series of papers \cite{SlowOCVp2}.

\subsection{Battery Capacity}

Table \ref{table:capdata} shows the estimated battery capacities for each experiment listed in Table \ref{table:data-collection-plan}. 
From this table, one can notice that the charge capacity is greater than the discharge capacity (except for the C/32 test which sits as an outlier) \cite{SlowOCVp2}. 

\begin{table*}[h!]
\caption{Battery capacity and internal resistance estimates calculated at different C-Rates.}
\begin{center}
\begin{tabular}{|c|c|c||c|c|c||c|c|c||c|c|}
\hline
Battery & C-Rate & $i$    & $Q_c$  & $Q_d$  & $Q_c - Q_d$ & $R_{oh}$    & $\hat R_0$  & $\hat R_\Omega$  & $V_d$ & $\bar h_2$ \\ 
        &        & (A)    & (Ah)   & (Ah)   & (Ah)        & (m$\Omega$) & (m$\Omega$) & (m$\Omega$) & (mV)  & (mV)     \\ \hline
D3209   & C/2    & 2      & 3.7860 & 3.9491 & -0.1631     & 35.3        & 17.39       & 12.23       & 70.63 & 35.89    \\ \hline
D3210   & C/2    & 2      & 3.8023 & 3.9645 & -0.1622     & 35.1        & 18.47       & 13.45       & 70.19 & 33.19    \\ \hline
D3211   & C/2    & 2      & 3.8268 & 3.9338 & -0.1070     & 38.2        & 16.99       & 11.82       & 76.45 & 42.39    \\ \hline
D3212   & C/2    & 2      & 3.8071 & 3.9168 & -0.1097     & 38.2        & 16.72       & 11.63       & 76.39 & 42.93    \\ \hline \hline
D3205   & C/4    & 1      & 3.9107 & 3.9567 & -0.0460     & 51.5        & 17.04       & 12.04       & 51.54 & 34.47    \\ \hline
D3206   & C/4    & 1      & 3.9218 & 3.9674 & -0.0456     & 52.3        & 16.71       & 11.50       & 52.28 & 35.60    \\ \hline
D3207   & C/4    & 1      & 3.9151 & 3.9615 & -0.0464     & 51          & 16.67       & 11.53       & 50.98 & 34.32    \\ \hline
D3208   & C/4    & 1      & 3.9076 & 3.9534 & -0.0458     & 50.4        & 16.67       & 11.67       & 50.41 & 33.72    \\ \hline \hline
D3201   & C/8    & 1      & 4.0421 & 4.0300 & 0.0121      & 75.5        & 18.01       & 12.88       & 37.73 & 28.73    \\ \hline
D3202   & C/8    & 0.5    & 4.0371 & 4.0246 & 0.0125      & 75.1        & 17.79       & 12.65       & 37.55 & 28.67    \\ \hline
D3203   & C/8    & 0.5    & 4.0389 & 4.0256 & 0.0133      & 75.3        & 18.48       & 13.35       & 37.66 & 28.43    \\ \hline 
D3204   & C/8    & 0.5    & 4.0394 & 4.0264 & 0.0130      & 75.1        & 18.12       & 12.86       & 37.57 & 28.51    \\ \hline \hline
D3201   & C/16   & 0.5    & 4.0731 & 4.0566 & 0.0165      & 125.8       & 18.01       & 12.04       & 31.46 & 26.96    \\ \hline
D3202   & C/16   & 0.25   & 4.0885 & 4.0722 & 0.0163      & 126.8       & 17.79       & 12.65       & 31.69 & 27.25    \\ \hline
D3203   & C/16   & 0.25   & 4.0781 & 4.0610 & 0.0171      & 126.8       & 18.48       & 13.35       & 31.71 & 27.10    \\ \hline
D3204   & C/16   & 0.25   & 4.0903 & 4.0741 & 0.0162      & 125.7       & 18.12       & 12.86       & 31.42 & 26.89    \\ \hline \hline
D3205   & C/32   & 0.125  & 4.0269 & 4.0369 & -0.0100     & 209.1       & 17.04       & 12.04       & 26.13 & 24.00    \\ \hline
D3206   & C/32   & 0.125  & 4.0446 & 4.0542 & -0.0096     & 209.4       & 16.71       & 11.50       & 26.17 & 24.09    \\ \hline
D3207   & C/32   & 0.125  & 4.0543 & 4.0649 & -0.0106     & 208.3       & 16.67       & 11.53       & 26.04 & 23.96    \\ \hline
D3208   & C/32   & 0.125  & 4.0359 & 4.0467 & -0.0108     & 207.7       & 16.67       & 11.67       & 25.96 & 23.87    \\ \hline \hline
D3209   & C/64   & 0.0625 & 4.1123 & 4.0974 & 0.0149      & 371         & 17.39       & 12.23       & 23.19 & 22.10    \\ \hline
D3210   & C/64   & 0.0625 & 4.1060 & 4.0898 & 0.0162      & 372.7       & 18.47       & 13.45       & 23.30 & 22.14    \\ \hline
D3211   & C/64   & 0.0625 & 4.1201 & 4.1052 & 0.0149      & 371.8       & 16.99       & 11.82       & 23.24 & 22.18    \\ \hline 
D3212   & C/64   & 0.0625 & 4.1077 & 4.0906 & 0.0171      & 371         & 16.72       & 11.63       & 23.19 & 22.14    \\ \hline \hline
D3213   & C/128  & 0.0312 & 4.1144 & 4.1103 & 0.0041      & 703.8       & 19.43       & 11.43       & 21.96 & 21.35    \\ \hline
D3214   & C/128  & 0.0312 & 4.1318 & 4.1283 & 0.0035      & 704.6       & 19.52       & 11.52       & 21.98 & 21.37    \\ \hline
D3215   & C/128  & 0.0312 & 4.1216 & 4.1196 & 0.0020      & 699.2       & 19.45       & 11.31       & 21.82 & 21.20    \\ \hline
D3216   & C/128  & 0.0312 & 4.1314 & 4.1293 & 0.0021      & 703.9       & 20.15       & 12.08       & 21.96 & 21.33    \\ \hline
\end{tabular}
\end{center}
\label{table:capdata}
\end{table*}

\begin{figure}[h!]
\begin{center}
{\includegraphics[width=0.8\columnwidth]{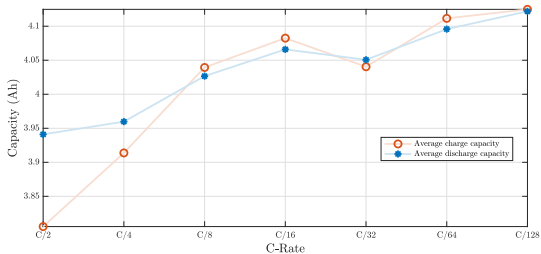}}
\caption{Charge and discharge capacities at different constant C-Rates of charging and discharging, respectively.}
\label{fig:QcQd}
\end{center}
\end{figure}

Figure \ref{fig:QcQd} shows the averaged charge capacity and averaged discharged capacity (averaging is done over all four cells);
the red curves denote the charge capacity and the blue curve denotes the discharge capacity. 
It can be noticed that the discharge capacity is significantly higher than the charge capacity at high currents.
Particularly, the charge and discharge capacities start to significantly deviate from one another below the C/4 rate. 
This is one indication that the assumptions made for the low-rate OCV test in \cite{SlowOCVp2} are not accurate when the current rate is high.

Table \ref{table:capdata} also shows the total resistance that is computed as, $\hat R_{oh} = \hat \bk_{\rm LS}(9)$.
Here, the notations introduced in part-2 of this series \cite{SlowOCVp2} are followed and $\hat \bk_{\rm LS}$ (OCV parameter vector) corresponds to the combined+3 model.

%

\subsection{Total Resistance}
\label{sec:R0estimates}

\begin{figure}[h!]
\begin{center}
\subfloat[][]
{\includegraphics[width=0.49\columnwidth]{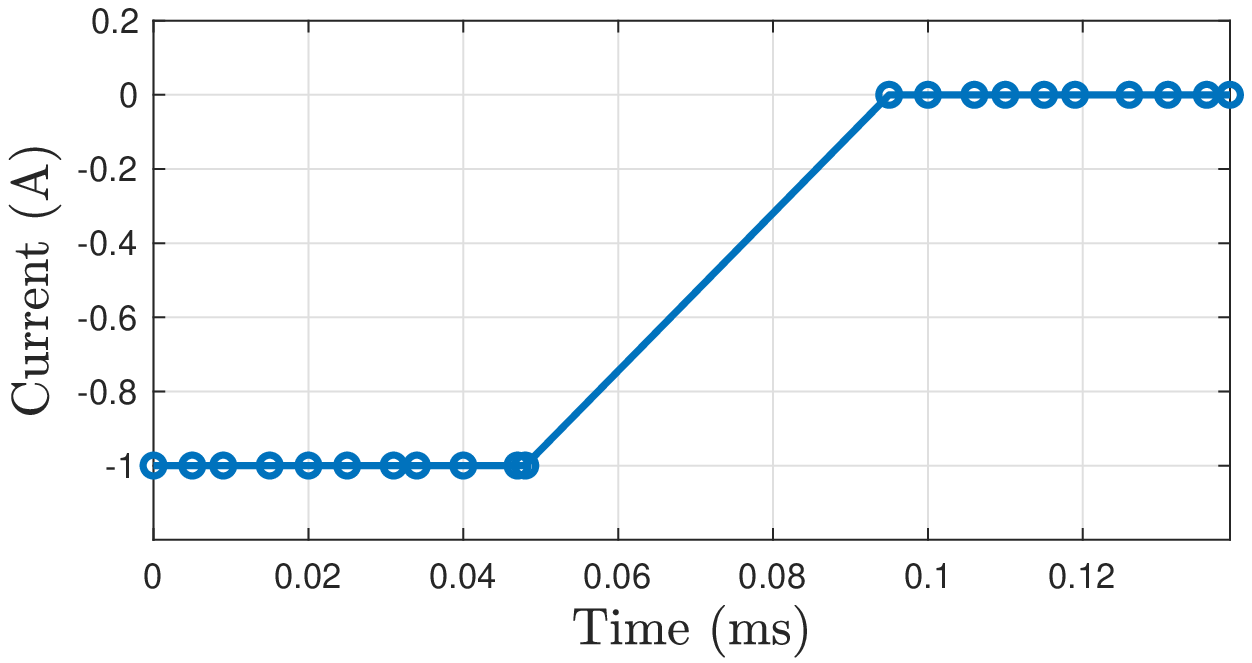}} 
\subfloat[][]
{\includegraphics[width= 0.49 \columnwidth]{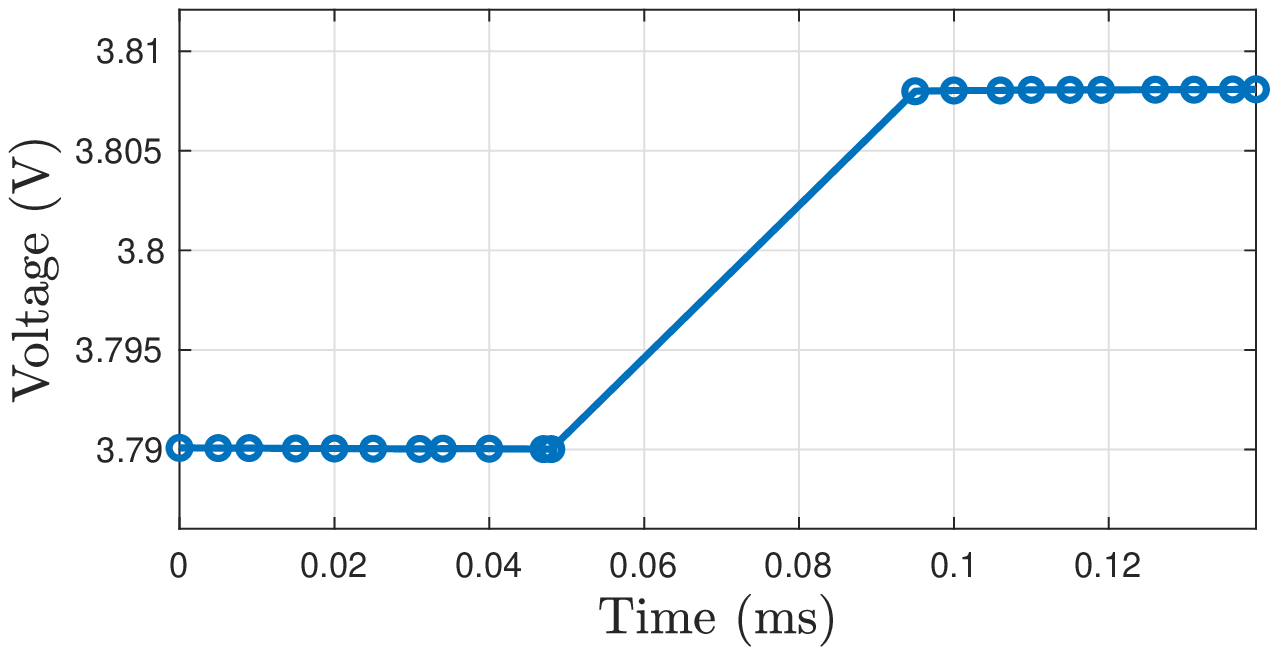}}
\caption{Current and voltage for resistance estimation (D3201)}
\label{fig:VIforR0est}
\end{center}
\end{figure}

The resistance of each battery was estimated using a pulse that consists of a discharge current of 1A for 50 ms followed by a rest of 50 ms; the resulting voltage across the battery is recorded for resistance estimation. In this case, since the batteries were closer to full after the low-rate OCV test, a current profile consisting only of discharge current was selected according to the suggested pulses in \cite{SlowOCVp2}. 
The sampling time is set as $\Delta = 5 \, {\rm ms}$, which is the lowest possible sampling time at the Arbin cycler.
Figure \ref{fig:VIforR0est} shows the voltage and current across a battery that would be used in resistance estimation using the least squares approach.  
More details on the resistance estimation pulses are presented in \cite{SlowOCVp2} and more details on the estimation approach and the performance bound can be found in \cite{pillai2022optimizing}. 
The time domain resistance estimation approach employed in this paper was evaluated based on a frequency domain approach in \cite{Wu2023Battery}.

In Figure \ref{fig:VIforR0est}, it can be seen that the battery tester takes significant time (about $45 \, {\rm ms}$) to switch from the discharge state to the rest state.

The battery resistance may vary with the SOC (and temperature). 
In this paper, all the experiments were conducted at room temperature and the batteries were full when the resistance estimation pulse (similar to the one shown in Figure \ref{fig:VIforR0est}) was applied. 
Also, according to the data reported in \cite{Wu2023Battery}, the change in internal resistance with respect to the SOC is insignificant for the tested batteries except at very low SOC regions of the battery.

Table \ref{table:capdata} shows the estimated resistance, $\hat R_0$, of each battery and compares them with those obtained by the Arbin battery testing system which utilizes a high-frequency signal at 1000 Hz to estimate the Ohmic resistance $R_\Omega.$
It is reported in \cite{pillai2022optimizing,barai2019comparison} that resistance estimated using high-frequency signals returns the Ohmic resistance $R_\Omega$ whereas that estimated using low-frequency signals return total resistance 
The observation always satisfy the expectation that $\hat R_0 >  \hat R_\Omega.$


%
%

\subsection{Hysteresis}
\label{sec:hysteresis}

From \cite{SlowOCVp2}, there are two ways to calculate the estimates of hysteresis:
\begin{align}
h_1(k) &= v(k) -  E(k) - i(k) \hat R_0  \label{eq:h1(k)}  \\
h_2(k) &= i(k) \hat R_h   \label{eq:h2(k)} 
\end{align}
where 
$v(k)$ is the measured terminal voltage of the battery, 
$i(k)$ is the current through the battery, and 
$E(k)$ denotes the estimate of OCV that is obtained based on the combined+3 model.

\begin{figure}[h!]
\begin{center}
\subfloat[][Terminal voltage.]
{\includegraphics[width=0.8\columnwidth]{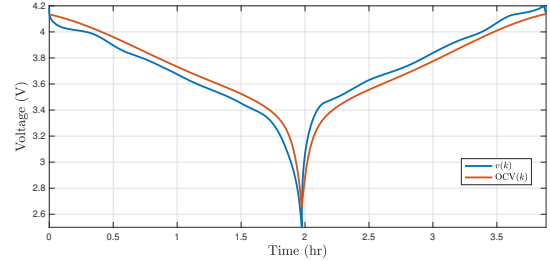}} \\
\subfloat[][Hysteresis voltage.]
{\includegraphics[width=0.8\columnwidth]{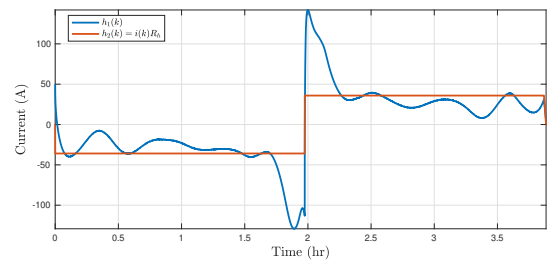}} 
\caption{Hysteresis at C/2 Rate Experiment (D3201).}
\label{fig:HysPlotModelOCV}
\end{center}
\end{figure}

Figure \ref{fig:HysPlotModelOCV} shows the hysteresis values computed according to \eqref{eq:h1(k)} and \eqref{eq:h2(k)}. 
First, the plot (a) in Figure \ref{fig:HysPlotModelOCV} shows the measured voltage $v(k)$ during the low-rate OCV experiment conducted at C/2 rate along with the corresponding OCV $V_o(k).$
The difference between $v(k)$ and $V_o(k)$ is accounted for by the hysteresis and relaxation effects of the battery. 

The plot (b) in Figure \ref{fig:HysPlotModelOCV} shows the hysteresis estimates $h_1(k)$ and $h_2(k)$ over time. 
Here, the hysteresis estimates $h_1(k)$ is instantaneous, as such, it varies with time, whereas $h_2(k)$ is an average measure of hysteresis and its magnitude remains constant during both charging and discharging; this is due to the assumption made in the empirical OCV modeling approach described in the preceding paper of this series \cite{SlowOCVp2}. 
It can be noticed that the hysteresis voltage is in the direction of the IR drop; i.e., it is negative during discharging and positive during charging. 
The effect of hysteresis will manifest differently at the battery terminals during charging and discharging; 
when charging, the terminal voltage will be higher than the combined OCV and relaxation effect; this would prevent the battery from getting fully charged;
when discharging, the terminal voltage will cause the battery to prematurely shut down before it can be fully used. 

\begin{figure}[h!]
\begin{center}
{\includegraphics[width=0.8\columnwidth]{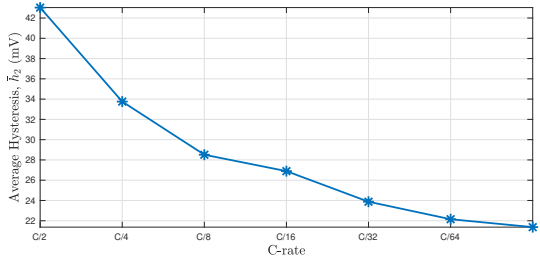}}
\caption{Average hysteresis values against the C-Rate.}
\label{fig:HysVcrate}
\end{center}
\end{figure}

The right-most column in Table \ref{table:capdata} shows the average hysteresis value that can be computed in two different ways as follows
\begin{align}
\bar h_2 = (C/N) \times \hat R_h = \sum_{k=1}^{m+n} \left| h_2(k) \right| 
\end{align}
From Table \ref{table:capdata}, one can notice that the hysteresis voltage increases with the current. 
Figure \ref{fig:HysVcrate} shows the average hysteresis values shown in Table \ref{table:capdata} for each current rate. 
The averaging is done over the four cells corresponding to each current rate.

\section{Observations and Conclusions}
\label{sec:conclusions}

The open circuit voltage (OCV) to state of charge (SOC) model plays an important role in battery management systems. 
The OCV-SOC curve (stored in the form of OCV-SOC parameters or as a table) is used for the accurate computation of the SOC of the battery.
Possible uncertainties in the OCV-SOC model: cell-to-cell variation, cycle-rate error and curve-fitting error, are computed and analyzed for low-rate OCV characterization tests done at seven C-Rates. 
For evaluating the different metrics, an existing OCV modeling approach known as pseudo-OCV modelling is applied. 
The following are some of the important observations from the analysis:
\begin{enumerate}
\item 
The mean and standard deviation due to the cell-to-cell variations was the highest at the C/2 rate. This width of uncertainty reduced as the C-Rate was reduced. 
\item
A non-symmetric error in cycle rate with respect to SOC was observed. The error increased when the battery was almost full/empty and also between ranges of 0.6 and 0.85. 
Further, the uncertainty corresponding to the cycle-rate variation was also found to increase with the increase in C-Rate. 
\item
The C-Rate error was found to increase with the current. The lower the C-Rate, the lesser the OCV modeling error and vice versa. 
\item
Curve fitting with respect to the different models has been significantly studied in the past \cite{pillai2022open, pattipati2014open}. It is known that the curve fitting error varies with the selected model. Of the three models selected in this paper, the curve fitting error decreased with the complexity of the model. 
\item
It is confirmed that the mean of all three types of errors was zero-mean with 95\% confidence.
\item
It is observed that the variance of all three types of errors varied with SOC. Particularly, the variance of the error was observed to be high at low/high SOC regions and approximately between SOC = 0.6 and SOC = 0.85. For a very accurate battery management system design, one may need to store these values against SOC to accurately account for the uncertainty.
\item
The hysteresis voltage was calculated across the four batteries for each C-Rate and it was observed that the voltage was directly proportional to the C-Rate. This also suggests selecting a low C-Rate for OCV modeling.
\end{enumerate}

The presented analysis on the performance of OCV modeling will be useful for battery researchers and manufacturers to select suitable parameters for OCV modeling in a way that the resulting uncertainty can be maintained within an expected level.

\bibliographystyle{ieeetr}
\bibliography{References}

\end{document}